\documentclass[aps,prb,twocolumn,superscriptaddress,showpacs,floatfix]{revtex4}

\usepackage{graphicx}
\graphicspath{{figs/}}
\begin{document}
\title{Molecular dynamics simulation for heat transport in thin diamond nanowires}
\author{Jin-Wu~Jiang}
    \altaffiliation{Electronic address: phyjj@nus.edu.sg}
    \affiliation{Department of Physics and Centre for Computational Science and Engineering,
             National University of Singapore, Singapore 117542, Republic of Singapore }
\author{Bing-Shen~Wang}
    \affiliation{State Key Laboratory of Semiconductor Superlattice and Microstructure and Institute of Semiconductor, Chinese Academy of Sciences, Beijing 100083, China}
\author{Jian-Sheng~Wang}
    \affiliation{Department of Physics and Centre for Computational Science and Engineering,
                 National University of Singapore, Singapore 117542, Republic of Singapore }

\date{\today}
\begin{abstract}
The phonon thermal conductivity in diamond nanowires (DNW) is studied by molecular dynamics simulation. It is found that the thermal conductivity in narrower DNW is lower and does not show obvious temperature dependence; a very small value (about 2.0 W/m/K) of thermal conductivity is observed in ultra-narrow DNW, which may be of potential applications in thermoelectric devices. These two phenomena are probably due to the dominant surface effect and phonon confinement effect in narrow DNW. Our simulation reveals a high anisotropy in the heat transport of DNW. Specifically, the thermal conductivity in DNW along [110] growth direction is about five times larger than that of [100] and [111] growth directions. The anisotropy is believed to root in the anisotropic group velocity for acoustic phonon modes in DNW along three different growth directions.
\end{abstract}

\pacs{65.80.-g, 62.23.Hj, 63.20.kg}
\maketitle

\pagebreak

\section{introduction}
Carbon-based materials have been attractive to both theoretical and experimental researchers for decades. From the bonding properties, these materials can be divided into two groups with different carbon-carbon bonds: sp2 bonding and sp3 bonding. The former includes the three-dimensional graphite, zero-dimensional fullerenes such as C$_{60}$, one-dimensional carbon nanotube, and two-dimensional graphene. These sp2 bonding materials have been investigated one by one and have achieved some industrial applications. As well known, the sp3 bonding diamond is the hardest material and of high thermal conductivity. Hsu {\it et~al.} have successfully synthesized diamond nanowires (DNW)\cite{HsuCH}, a new carbon-based material with sp3 bonding inner the system. DNW in the experiment are grown by a bottom-up technology, and the obtained samples are about 60-90 nm in diameter along the [110] growth direction. Babinec {\it et.al} have fabricated 200-nm-diameter DNW by top-down method, and used the DNW samples as a single-photon source for the photonic and quantum information processing.\cite{Babinec} After the gold rush for diamond, fullerences, carbon nanotube, and graphene, the DNW is a competitive candidate of carbon-based materials for future novel applications. The DNW may inherit various advanced properties from diamond and gain new characters from its nanowire configuration of tunable surface to volume ratio (SVR).

Prior to the experiment, in 2003, Barnard {\it et~al.} have done a series of theoretical works to investigate the stability and possible phase transition for the structure of DNW.\cite{Barnard1,Barnard2,Barnard3,Barnard5,Shenderova,Barnard6} The {\it ab intio} calculations show that the structure relaxation of DNW depends on both the surface morphology and the growth direction\cite{Barnard1}. The effect of the boron and nitrogen dopants on the structure stability of the DNW was also examined by the same group\cite{Barnard2}. They also found that the energy band gap of DNW is reduced considerably due to the effect of surface states\cite{Barnard4}. In 2008, Tanskanen {\it et.al} carried out quantum chemical calculations to study the the structural, electronic, and mechanical properties of DNW.\cite{Tanskanen} Since 2004, there is quite few theoretical works on the DNW, as it was a challenge to grow the DNW for experimentalists at that time. As now the DNW samples have been successfully prepared, it is an urgent task to pursue theoretical studies on important properties of the DNW. The thermal transport is one of the very important phenomena in the quasi-one-dimensional nanowire structures, where the phonon spectrum modification and boundary scattering are important.\cite{Khitun,Zou} Some possible practical applications of the phonon confinement effect has also been found.\cite{Fonoberov}

In this paper, we investigate the thermal conductivity in the DNW by molecular dynamics simulation using the Tersoff\cite{Tersoff} and Brenner\cite{Brenner} empirical inter-atomic potentials. We show that the thermal conductivity of thicker DNW decreases with increasing temperature due to phonon phonon scattering (PPS), yet does not depend on temperature in narrower DNW where surface scattering is important. Our simulation discloses a highly anisotropic heat transport in DNW: thermal conductivity in DNW along [110] growth direction is about five times higher than that of [100] and [111] growth directions. By performing lattice dynamics analysis for DNW, we believe that this anisotropy is originating from anisotropic group velocity of acoustic phonon modes in DNW along three different growth directions.

\begin{figure}[htpb]
  \begin{center}
    \scalebox{1.0}[1.0]{\includegraphics[width=8cm]{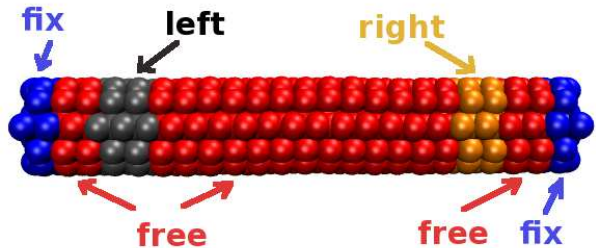}}
  \end{center}
  \caption{(Color online) Configuration of DNW[110](60,10). Two ends (blue online) are fixed. The high/low temperature controlled regions are on left and right. Other atoms (red online) are free during simulation.}
  \label{fig_cfg}
\end{figure}
\section{structure of DNW and simulation details}
The DNW shown in Fig.~\ref{fig_cfg} is denoted as DNW[110](60,10), indicating that DNW is growing along [110] direction with length $L=60$~{\AA} and diameter $D=10$~{\AA}. We adopt this notation in the following. The DNW is cut from the bulk diamond crystal by using a virtual cylinder with structure parameters ($L$, $D$)\cite{VoT}, so that the cross section is always a circle and the diameter is well defined to be the diameter of the virtual cylinder. We applied free boundary condition for the surface of the nanowires, i.e atoms on the surface have free bonds. In thermal transport, these free bonds introduce some roughness to the surface of the nanowires and they can reduce thermal conductivity.\cite{Ouyang,Markussen} The macroscopic word `roughness' is always borrowed to describe the microscopic structure of the nano-devices surface. The surface roughness can be introduced to manipulate the thermal conductivity through generating some vacancies on the surface which produces some free bonds on the surface. In this sense, the free boundary condition on the surface is actually equivalent to a rough surface. An experimental group has taken advantage of the rough surface to enhance the thermoelectric performance of silicon nanowires.\cite{Hochbaum} Hence, the free boundary condition on the surface is of practical importance, so we prefer to use it in our simulation. In the molecular dynamics simulation, the free bonds of surface atoms are taken care naturally by the bond-order Tersoff potential. The local environment of surface atoms is different from inner atoms, and this difference is reflected by the bond-order parameters of the Tersoff potential. Both ends are fixed during MD simulation. The heat energy is pumped into the DNW through the temperature-controlled region on the left (black online) with thermal current $J_{L}$ and will flow out of the DNW through the temperature-controlled region on the right (yellow online) with current $J_{R}$. At steady state, energy conservation requires that $J_{L}=-J_{R}$. Using this relation, the thermal current flowing through the DNW can be obtained by $J=(J_{L}-J_{R})/2$, with $dJ=J_{L}+J_{R}$ as an estimated of error. All other atoms (red online) vibrate freely. It is important to keep a free segment between fixed end and the temperature-controlled region to avoid energy accumulations and big boundary temperature jumps.\cite{JiangJW}

\begin{figure}[htpb]
  \begin{center}
    \scalebox{0.9}[0.9]{\includegraphics[width=8cm]{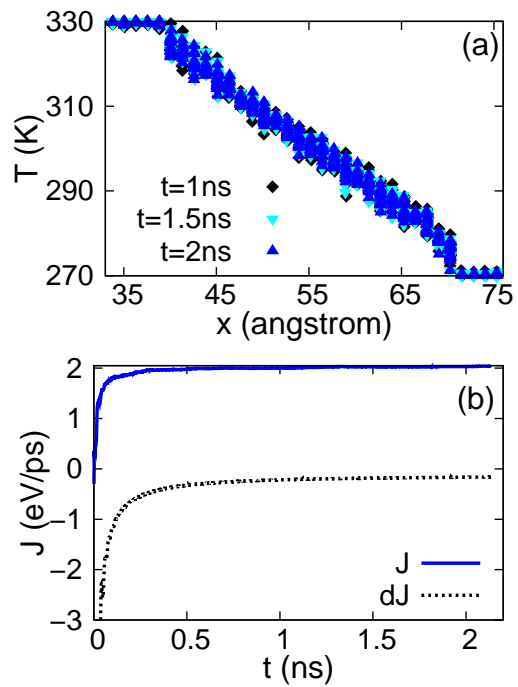}}
  \end{center}
  \caption{(Color online) Thermal quantities in DNW[110](60,10) at 300K. (a). Temperature profile at three different simulation time. (b). Thermal current across the DNW.}
  \label{fig_time}
\end{figure}
The N\'ose-Hoover\cite{Nose, Hoover} thermostat is employed to maintain constant temperature in the left and right temperature-controlled regions. The force between carbon atoms in the Newton equation is described by the commonly used Tersoff\cite{Tersoff} and Brenner\cite{Brenner} empirical inter-atomic potentials. The Newton equations are solved by velocity Verlet algorithm with an integration time step of 1.0 fs. A total simulation time is typically around 2.0 ns and will be sufficiently extended to guarantee the achievement of steady state (if necessary). Fig.~\ref{fig_time} exhibits that the current difference $dJ$ is almost zero at 1.0 ns and the thermal current has already reached a saturate value. There is no difference between temperature profiles at 2.0 ns and 1.0 ns. These facts show that the simulation time is long enough.

\section{results and discussion}
\begin{figure}[htpb]
  \begin{center}
    \scalebox{0.9}[0.9]{\includegraphics[width=8cm]{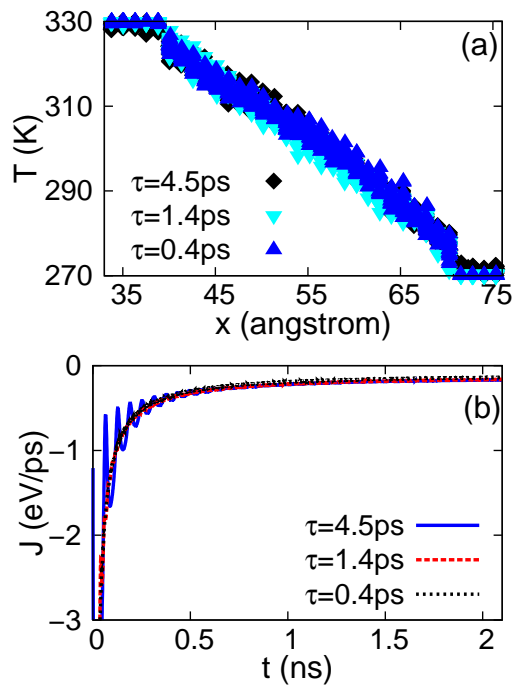}}
  \end{center}
  \caption{(Color online) Heat transport in DNW[110](60,10) at 300K with different relaxation time of heat bath. (a). Temperature profile. (b). Thermal current.}
  \label{fig_tau}
\end{figure}
In the N\'ose-Hoover thermostat, the effective relaxation time $\tau$ determines the temperature fluctuation for temperature-controlled regions and the strength of connection between heat bath and DNW. A small $\tau$ yields small temperature fluctuation due to strong interaction between heat bath and system, which facilitates energy injection into DNW. Thus the DNW can arrive at steady state faster and will save much simulation time. On the other side, large $\tau$ corresponds to a weak heat bath with more fluctuations as shown in Fig.~\ref{fig_tau}(b), resulting in longer simulation time. From Fig.~\ref{fig_tau}(a), all relaxation time $\tau$=4.5, 1.4, and 0.4 ps accounts for almost the same temperature profile. We will use $\tau=1.4$ ps in the following, so that the DNW can reach steady state quickly and without introducing much thermal noise to system.

\begin{figure}[htpb]
  \begin{center}
    \scalebox{0.9}[0.9]{\includegraphics[width=8cm]{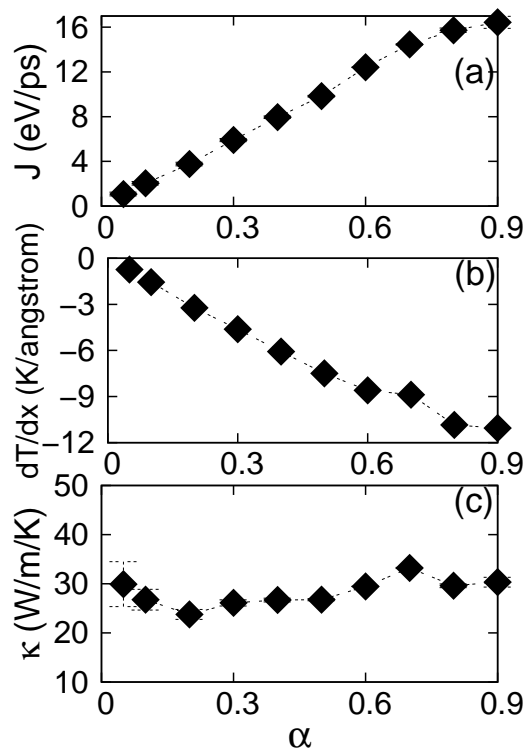}}
  \end{center}
  \caption{Heat transport in DNW[110](60,10) at 300K with different temperature ratio $\alpha$. (a). The thermal current. (b). The temperature gradient. (c). Thermal conductivity.}
  \label{fig_Tratio}
\end{figure}

Following the Fourier law, the thermal conductivity $\kappa$ is obtained as a ratio of current to temperature gradient: $\kappa=-J/(dT/dx)$. The temperature gradient is calculated from linear fitting to the temperature profile in the free region between left and right heat baths as shown in Fig.~\ref{fig_cfg}. To drive the thermal current across the DNW, the left/right temperature-controlled regions in Fig.~\ref{fig_cfg} are put in heat baths with $T_{L/R}=(1\pm\alpha)T$. $T$ is the average temperature of the DNW and $\alpha$ is the ratio of temperature difference. If $\alpha$ is large, the big temperature difference can drive a large thermal current and set up a big temperature gradient. As a result, the thermal conductivity will be almost unaffected provided that the Fourier law still valid. Fig.~\ref{fig_Tratio} confirm this result. Panels (a) and (b) show that the thermal current and the temperature gradient increase linearly with increasing $\alpha$. The resulted thermal conductivity shown in panel (c) only display small variation with $\alpha$ increasing from 0.05 to 0.9. We will keep $\alpha=0.1$ in following calculations.

\begin{figure}[htpb]
  \begin{center}
    \scalebox{0.9}[0.9]{\includegraphics[width=8cm]{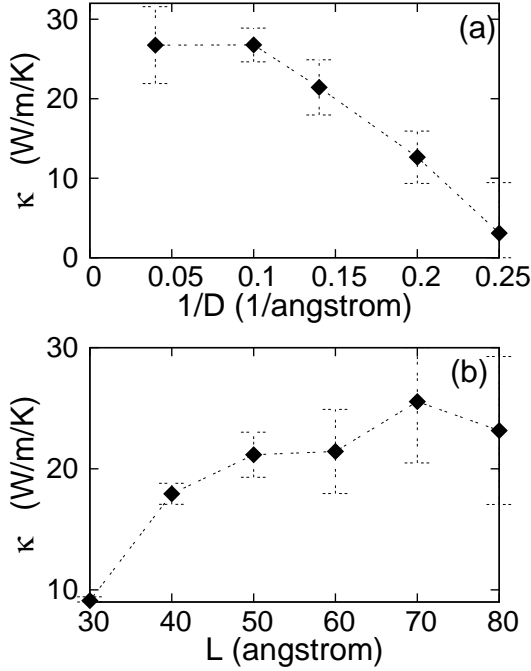}}
  \end{center}
  \caption{Thermal conductivity at 300 K in DNW[110] with different size (a). $L=60$~{\AA} and (b). $D=7$~{\AA}.}
  \label{fig_size}
\end{figure}

The thermal conductivity at room temperature in DNW[110] of different length and diameter are shown in Fig.~\ref{fig_size}. $\kappa$ of DNW[110] with length as 60~{\AA} and different diameters are displayed in panel (a). Only a small variation in thermal conductivity is observed between thick DNW. The thermal conductivity decreases with decreasing diameter, due to the enhancement of surface scattering in narrower DNW. A very low thermal conductivity of about 2.0 W/m/K is obtained for the ultra-narrow DNW. The poor capability of heat transport may find some use in the thermoelectric applications where lower thermal conductivity is desired. However, we do not observe the increase of thermal conductivity with decreasing diameter, which was discovered in silicon nanowires by Ponomareva {\it et~al.}\cite{Ponomareva}. It should be note, however, from panel (a), the value of $\kappa$ almost does not change for the two DNW of largest diameters in the figure, but it is still difficult to predict the value of thermal conductivity for experimental DNW samples with about an order of magnitude larger in diameter. Panel (b) shows $\kappa$ in DNW[110] with same diameter 7~{\AA}. The thermal conductivity increases with the increase of length below 40~{\AA}, However, no obvious change is found for $\kappa$ above 40~{\AA}, which indicates a diffusive thermal transport instead of ballistic transport. It is quite reasonable to observe diffusive thermal transport in DNW, since we have successfully obtained the linear temperature profile from the direct MD simulation as shown in Fig.~\ref{fig_time}~(a) and Fig.~\ref{fig_tau}(a). As a result of surface effect, the value of thermal conductivity in DNW is much smaller than that of the diamond (about 2000 W/m/K)\cite{SukhadolauAV}. Similar phenomenon has been found in silicon nanowires\cite{LiD}.

\begin{figure}[htpb]
  \begin{center}
    \scalebox{1.0}[1.0]{\includegraphics[width=8cm]{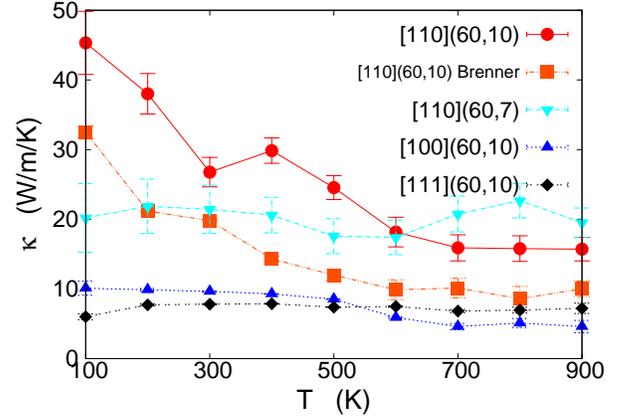}}
  \end{center}
  \caption{(Color online) Thermal conductivity v.s temperature in different DNW with different types of interaction.}
  \label{fig_tem}
\end{figure}

Fig.~\ref{fig_tem} shows the temperature dependence for thermal conductivity in different DNW. For DNW[110](60,10), we have applied both Tersoff (circle, red online) and Brenner (square, red online) potentials to describe the interaction between carbon atoms. Both curves show a similar trend that $\kappa$ decreases with increasing temperature in the whole temperature range. This result manifests the importance of the phonon-phonon scattering in DNW[110](60,10). The PPS happens inside the system and is stronger at higher temperature\cite{Holland}. The relaxation time of phonons corresponding to normal PPS at high temperature is inverse proportional to temperature $T$, leading to lower $\kappa$ at higher $T$. Compared with Tersoff potential, the Brenner potential gives an obviously smaller value for $\kappa$ in whole temperature range. Both potentials are multi-body interaction and include nonlinear interactions. They have been widely applied to describe the carbon-carbon interactions and the bonding process. It is not easy to compare the difference in the nonlinear interaction between these two potentials. However, we can extract some valuable information on the nonlinear properties of these two potentials from the thermal conductivity of these two potentials in Fig.~\ref{fig_tem}. With increasing $T$, two curves in the figure decline at almost the same speed, which implies that the nonlinear interactions in two potentials are more or less the same. Actually, the linear part of these two potentials has been compared, and the Brenner potential gives weaker interaction than Tersoff potential for carbon systems.
 The frequency of optical phonon modes from Brenner potential is overall smaller than that of the Tersoff potential. This difference is responsible for the lower thermal conductivity from Brenner potential than Tersoff potential in Fig.~\ref{fig_tem}. Experimental works are in need to judge which potential is more suitable to be applied in the study of thermal conductivity of DNW. We use Tersoff potential in all other calculations of this paper.

\begin{figure}[htpb]
  \begin{center}
    \scalebox{1.0}[1.0]{\includegraphics[width=8cm]{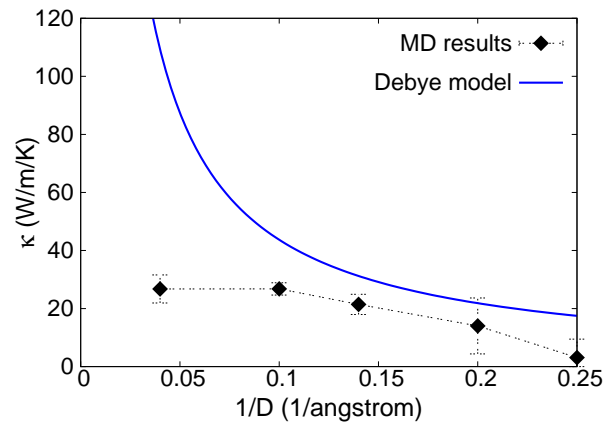}}
  \end{center}
  \caption{(Color online) Thermal conductivity v.s 1/diameter at 300 K in DNW[110] with length 60~{\AA}. The MD simulation results (black filled square) are compared with simple Debye thermal conductivity (blue line).}
  \label{fig_debye}
\end{figure}

It is quite interesting that the thermal conductivity of DNW[110](60,7) in Fig.~\ref{fig_tem} does not depend on the temperature. This is different from the thicker DNW[110](60,10), where the thermal conductivity is determined by the PPS as discussed above. The thermal conductivity is not sensitive to temperature probably because of the dominant boundary scattering and phonon confinement effects in thin DNW. In the thermal transport of nano-materials such as DNW, the surface effect is another important mechanism due to confinement effect of the DNW. The PPS has dominant effect in bulk systems or nanowires with low SVR, while the surface effect determines the thermal conductivity in ultra narrow DNW with high SVR\cite{MartinP,LiD}. The surface scattering can be described in various forms. A more widely used formula for the relaxation rate of surface scattering can be found from Ziman's book:\cite{Ziman}
\begin{eqnarray}
\frac{1}{\tau_{\rm bs}} & = & \frac{v_{q\sigma}}{L}\times\frac{1-p}{1+p},
\label{eq_bslifetime}
\end{eqnarray}
where $v_{q\sigma}$ is the phonon velocity of mode $(q,\sigma)$, and $L$ is the sample size. $\tau_{\rm bs}$ is the corresponding life time and $p$ is the spectacular parameter. The thermal conductivity can be obtained from the kinetic theory:
\begin{eqnarray}
\kappa = \frac{1}{V} \sum_{q,\sigma} \tau_{q\sigma}C_{\rm ph}(\omega)v_{q,\sigma}^{2},
\label{eq_conductivity}
\end{eqnarray}
where $V$ is the volumn and $C_{\rm ph}(\omega)$ is the phonon heat capacity. The life time due to boundary scattering is temperature independent, so the temperature dependence for the thermal conductivity is the same as the heat capacity; i.e, $\kappa$ increases with increasing $T$. The temperature dependence of $\kappa$ is determined by the competition between the PPS and surface scattering. In thinner diamond nanowires, the surface scattering is also important; thus the interaction between these two mechanisms leads to much smooth temperature dependence for $\kappa$ and even results in the absence of $1/T$ character. That is the origin of temperature insensitive thermal conductivity in DNW[110](60,7). We have applied free boundary condition for the surface of the nanowire, so atoms on the surface have less than four neighboring carbon atoms. This particular surface configuration is taken care by the bond-order Tersoff potential which we have applied to describe the interatomic interactions. During the molecular dynamics simulation, the Newton equations of all carbon atoms are numerically solved, including the surface atoms. The movement of the surface atoms is different from the inner atoms, because of different local environment in the Tersoff potential. Actually, there are some surface phonon modes on the surface, which contribute to the surface density of state as indicated by Fig.~\ref{fig_debye}. As a result, the surface atoms introduce some roughness effect to the thermal transport, which eventually leads to the surface scattering. The bond-order Tersoff potential includes the nonlinear interactions such as $u^{3}$ in the real space with $u$ as the vibrational displacement. A Fourier transformation of this term leads to the three-phonon scattering process. So the PPS is also automatically included in the molecular dynamics simulation through the Tersoff potential. We can not distinguish surface scattering and PPS in the molecular dynamics simulation in the real space, because these two scatterings are differentiated in the eigen mode space (the reciprocal space). The thermal conductivity depends strongly on the details of the surface configuration. From molecular dynamics simulation, different surface configuration will result in different local environment for surface atoms, which is reflected through the bond-order term of the Tersoff potential. This will directly affect the heat transport in the diamond nanowires. From kinetic theory, different surface configuration will result in different value of the spectacular parameter $p$ in Eq.~(\ref{eq_bslifetime}). It also has straight-forward effect on the value of $\kappa$. So there will be some difference if the surface boundary condition is not free as used in our work.

Eq.~(\ref{eq_bslifetime}) and (\ref{eq_conductivity}) can give a more rigorous description for the surface scattering, yet there is an adjustable parameter $p$. This parameter need to be fitted to the experimental results.\cite{NikaDL} As there is no experiment here, instead, we calculate the thermal conductivity $\kappa$ by the Debye model with formula:
\begin{eqnarray}
\kappa=\frac{1}{V_{a}}\sum_{q,\sigma}C(\omega_{q,\sigma})v_{q,\sigma}l,
\label{eq_debye}
\end{eqnarray}
where $V_{a}$ is the volumn per atom. $C_{q,\sigma}$ and $v_{q,\sigma}$ are the heat capacity and velocity of phonon mode $(q,\sigma)$. $q$ is the wave vector and $\sigma$ is the branch index for the phonon spectrum.
 In the quasi-one-dimensional diamond nanowire, the four acoustic phonon branches dominate the thermal conductivity, so only these four acoustic phonon branches are considered. The velocities of these four acoustic branches are obtained through $v=d\omega/dq$ from their spectrum. The mean free path $l$ is taken to be the diameter of the nanowire. It corresponds to strong roughness of the surface, indicating the absorption of phonon as soon as the phonon reaches the surface. This condition is reasonable for diamond nanowires of small diameter where strong boundary scattering has been found by the MD simulation. Fig.~\ref{fig_debye} shows the thermal conductivity from our MD simulation (black filled square) and this simple Debye model (blue line). The diamond nanowires studied here are DNW[110] with length 60~{\AA} and different diameters. The four acoustic phonon velocities are 21.3, 10.8, 10.5, and 9.3 km/s. The Debye model is initially established in the bulk material; however, it is quite interesting that this Debye thermal conductivity is close to the MD results for nanowires of small diameter, and $\kappa$ decreases with decreasing diameter, considering its extraordinary simplicity. In the large diameter region, the Debye model overestimates the value of thermal conductivity, because the phonon-phonon scattering is completely ignored in this model. It should be note that the velocities for phonon modes near the boundary of Brillouin zone become smaller, which is another reason for the overestimation of $\kappa$ by Debye model.

\begin{figure}[htpb]
  \begin{center}
    \scalebox{1.0}[1.0]{\includegraphics[width=8cm]{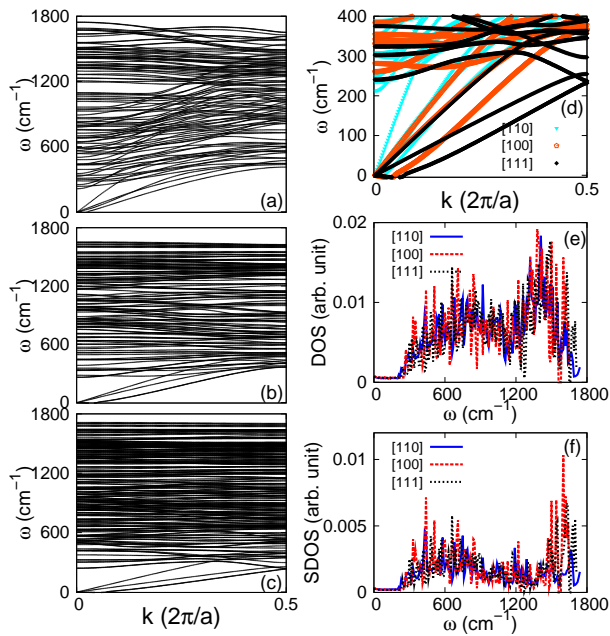}}
  \end{center}
  \caption{(Color online) Phonon modes in DNW with size (60,10)~{\AA} in three growth directions. (a)-(c) are phonon dispersion for DNW in [110], [100], and [111] growth directions. (d). The difference of phonon dispersion between three growth directions in low frequency region. (e). The DOS of phonon modes. (f). The DOS projected onto surface atoms.}
  \label{fig_phonon}
\end{figure}
In Fig.~\ref{fig_tem}, we compare the thermal conductivity of DNW in three different growth directions [100], [110], and [111]. These DNW are of the same size (60,10). A highly anisotropic thermal conductivity is observed: $\kappa$ in DNW[110] is about five times larger than the other two growth directions. To reveal the origin for this big difference in thermal conductivity, we study phonon modes in DNW along three growth directions as shown in Fig.~\ref{fig_phonon}. In panels (a)-(c), the wave vector $k$ is in the dimension of reciprocal lattice of the one-dimensional Brillouin zone space, where $a=2.51$, 3.56, and 6.16~{\AA} is the lattice constant for DNW along [110], [100], and [111] growth directions. We have shown all phonon modes in half of the full Brillouin zone for three DNW. The group velocity of the acoustic phonon modes can influence the thermal conductivity in three aspects. Firstly, a larger group velocity directly leads to higher thermal conductivity, which is proportional to phonon group velocity and relaxation time\cite{Holland}. Secondly, larger group velocity will result in smaller possibility of Umklapp PPS (which can reduce thermal conductivity) for these acoustic phonon modes in low temperature region. This is because they have smaller wave vectors for the same frequency, thus the possibility of Umklapp PPS is reduced. This effect further enhances the thermal conductivity. The importance of Umklapp PPS on thermal conductivity was also found in graphene\cite{NikaDL}. Thirdly, with increasing wave vector $k$, the dispersion curve for acoustic phonon mode of larger group velocity will increase and resonate with the dispersion curve of optical phonon modes more quickly. The resonance between acoustic and optical phonon modes leads to strong PPS, thus reducing the thermal conductivity in high temperature region. In one word, for systems with acoustic phonon modes of larger group velocity, the first two mechanisms will enhance the thermal conductivity, while the third mechanism reduce the thermal conductivity and leads to obvious temperature dependence of the thermal conductivity. The third mechanism will compete with the first two mechanisms and it will be more important at high temperatures. For DNW, there are four acoustic phonon modes in the low frequency region: one longitudinal mode, two degenerated transverse modes and a twisting mode. As the low-frequency phonon modes are important, we compare the four acoustic phonon dispersions in Fig.~\ref{fig_tem}~(d). Obviously, the group velocity of acoustic phonon modes are in the order of [110]$>$[100]$>$[111]. The largest group velocity of acoustic phonon modes in DNW[110] is due to its strongest Youngs moduli (22.8 GPa) that is about $50\%$ larger than the other two directions. The strong Youngs moduli may have relation to the redistribution of the elastic deformations during the relaxation of the surface atoms in the DNW, which happens in the acoustically mismatched nanowires as discussed by Pokatilov {\it et.al}.\cite{PokatilovPRB,PokatilovJSM} As discussed above, the largest group velocity in DNW[110] leads to highest thermal conductivity in DNW[110] and obvious temperature dependence for $\kappa$ shown in Fig.~\ref{fig_tem}. Due to the same reason, the thermal conductivity is much smaller in DNW[100] and only depends on temperature above 500 K. The smallest group velocity in DNW[111] results in lowest thermal conductivity and temperature insensitive $\kappa$ in whole temperature range.

Besides the low frequency acoustic phonon modes, we should examine the high frequency optical modes, which may play some role in the anisotropy of thermal conductivity even though these modes have limited contribution to $\kappa$ due to their very small group velocity. We calculate the total density of state (DOS) for phonon modes in three growth directions as compared in panel (e). The major difference in DOS is found in high frequency region around 1600 cm$^{-1}$. The DOS is obviously larger in [100] and [111] growth directions compared with [110] growth direction. At first glance, this should lead to higher thermal conductivity in these two directions. However, we further find that the surface modes are the major component of the high frequency phonon modes in DNW[100] and DNW[111]. This fact is disclosed by the surface DOS (SDOS), the projection of DOS onto surface atoms, as shown in panel (f). The SDOS is much larger in DNW[100] and DNW[111], resulting in very small contribution from these high frequency optical phonon modes to thermal conductivity in DNW[100] and DNW[111].

\section{conclusions}
In conclusion, by carrying out molecular dynamics simulation, we have investigated the thermal conductivity in the DNW with the widely used Tersoff and Brenner empirical inter-atomic potentials. We find that the thermal conductivity in thicker DNW is lower at higher temperatures, while $\kappa$ in narrower DNW is insensitive to temperature. The thermal conductivity shows decreasing behavior with decreasing diameter of DNW and $\kappa\approx 2.0$~{W/m/K} in ultra-narrow DNW which benefits its thermoelectric applications. We demonstrate that the thermal conductivity in DNW[110] is about five times larger than that of DNW[100] and DNW[111] of the same size. The obtained results are interpreted by PPS and the anisotropic group velocity of acoustic phonon modes. Diffusive transport is reached for DNW with lengths larger than 50~{\AA} and diameter of 7~{\AA}.

\textbf{Acknowledgements} We thank the anonymous referee for the suggestion of Debye model. The work is supported in part by a Faculty Research Grant of R-144-000-257-112 of National University of Singapore.


\begin{thebibliography}{}
\bibitem{HsuCH} C.-H. Hsu, S. G. Cloutier, S. Palefsky, and J. Xu, Nano. Lett. \text{10}, 1021 (2010).

\bibitem{Babinec} T. M. Babinec, B. J. M. Hausmann, M. Khan, Y. Zhang, J. R. Maze, P. R. Hemmer, and M. Loncar, Nature Nanotechnology \textbf{5}, 195 (2010).

\bibitem{Barnard1} A. S. Barnard, S. P. Russo, and I. K. Snook, Nano. Lett. \textbf{3}, 1323 (2003).

\bibitem{Barnard3} A. S. Barnard, S. P. Russo, and I. K. Snook, Philosophical Magazine, \textbf{83}, 2311 (2003).

\bibitem{Barnard5} A. S. Barnard, S. P. Russo, and I. K. Snook, Surface Science \textbf{538}, 204 (2003).

\bibitem{Shenderova} O. Shenderova, D. Brenner, and R. S. Ruoff, Nano. Lett. \textbf{3}, 805 (2003).

\bibitem{Barnard6} A. S. Barnard and I. K. Snook, J. Chem. Phys. \textbf{20}, 3817 (2004).

\bibitem{Barnard2} A. S. Barnard, S. P. Russo, and I. K. Snook, Philosophical Magazine, \textbf{83}, 2301 (2003).

\bibitem{Barnard4} A. S. Barnard, S. P. Russo, and I. K. Snook, Phys. Rev. B \textbf{68}, 235407 (2003).

\bibitem{Tersoff} J. Tersoff, Phys. Rev. B \textbf{38}, 9902 (1988).

\bibitem{Brenner} D. W. Brenner , O. A. Shenderova , J. A. Harrison, S. J. Stuart, B. Ni and S. B. Sinnott, J. Phys.:Condens. Matter \textbf{14}, 783 (2002).

\bibitem{VoT} T. Vo, A. J. Williamson, and G. Galli, Phys. Rev. B \textbf{74}, 045116 (2006).

\bibitem{JiangJW} J. W. Jiang, J. Chen, J.-S. Wang, and B. Li, Phys. Rev. B \textbf{80}, 052301 (2009).

\bibitem{Nose} S. N\'ose, J. Chem. Phys. \textbf{81}, 511 (1984).

\bibitem{Hoover} W. G. Hoover, Phys. Rev. A, \textbf{31}, 1695 (1985).

\bibitem{Ponomareva} I. Ponomareva, D. Srivastava, and M. Menon, Nano. Lett. \textbf{7}, 1155 (2007).

\bibitem{SukhadolauAV} A. V. Sukhadolau, E. V. Ivakin, V. G. Ralchenko, A. V. Khomich, A. V. Vlasov, and A. F. Popovich, Diamond Relat. Mater. \textbf{14}, 589 (2005).

\bibitem{LiD} D. Li, Y. Wu, P. Kim, L. Shi, P. Yang, and A. Majumdar, Appl. Phys. Lett. \textbf{83}, 2934 (2003).

\bibitem{Holland} M. G. Holland, Phys. Rev. \textbf{132}, 2461 (1963).

\bibitem{MartinP} P. Martin, Z. Aksamija, E. Pop, and U. Ravaioli, Phys. Rev. Lett. \textbf{102}, 125503 (2009).

\bibitem{NikaDL} D. L. Nika, E. P. Pokatilov, A. S. Askerov, and A. A. Balandin, Phys. Rev. B \textbf{79}, 155413 (2009).

\bibitem{Ziman} J. M. Ziman, \textit{Electrons and Phonons} (Clarendon Press, Oxford, 1960).

\bibitem{Khitun} A. Khitun, A. Balandin, and K. L. Wang, J. Superlattices Microstruct. \textbf{26}, 181 (1999).

\bibitem{Zou} J. Zou and A. Balandin, J. Appl. Phys. \textbf{89}, 2932 (2001).

\bibitem{Fonoberov} V. A. Fonoberov and A. A. Balandin, Nano. Lett. \textbf{6}, 2442 (2006).

\bibitem{PokatilovPRB} E. P. Pokatilov, D. L. Nika, and A. A. Balandin, Phys. Rev. B \textbf{72}, 113311 (2005).

\bibitem{PokatilovJSM} E. P. Pokatilov, D. L. Nika, and A. A. Balandin, J. Superlattices Microstruct. \textbf{38}, 168 (2005).

\bibitem{Tanskanen} J. T. Tanskanen, M. Linnolahti, A. J. Karttunen, and T. A. Pakkanen, J. Phys. Chem. C \textbf{112}, 11122 (2008).

\bibitem{Ouyang} Y. Ouyang and J. Guo, Appl. Phys. Lett. \textbf{94}, 263107 (2009).

\bibitem{Markussen} T. Markussen, A.-P. Jauho, and M. Brandbyge, Phys. Rev. B \textbf{79}, 035415 (2009).

\bibitem{Hochbaum} A. I. Hochbaum, R. Chen, R. D. Delgado, W. Liang, E. C. Garnett, M. Najarian, A. Majumdar, and P. Yang, Nature \textbf{451}, 163 (2008).

\end{thebibliography}
\end{document}